\documentclass[nohyper,11pt,letterpaper]{JHEP3}
\pdfoutput=1
\usepackage{amssymb,amsmath}
\usepackage{epsfig}
\usepackage{color}
\usepackage{datetime}

\newcommand{\BM}[1]{{\mbox{\boldmath $#1$}}}

\newcommand{\be}{\begin{equation}}
\newcommand{\ee}{\end{equation}}
\newcommand{\ba}{\begin{eqnarray}}
\newcommand{\ea}{\end{eqnarray}}
\newcommand{\hh}{\, ,\hspace{0.5cm}}
\newcommand{\hhh}{\, ,\hspace{0.2cm}}

\newcommand{\lap}{\bigtriangleup}

\newcommand{\n}[1]{\label{#1}}

\newcommand{\inds}[1]{{\scriptscriptstyle #1}}

\newcommand{\llan}{\langle\!\langle}
\newcommand{\rran}{\rangle\!\rangle }

\title{Spherical collapse of small masses in the ghost-free gravity}

\author{Valeri P. Frolov $^{(a)}$\thanks{E-mail:
vfrolov@ualberta.ca},
Andrei Zelnikov  $^{(a)}$\thanks{E-mail: zelnikov@ualberta.ca}\,  and
Tib\'{e}rio de Paula Netto $^{(a,b)}$\thanks{E-mail: depaulan@ualberta.ca }\\
$^{(a)}$ Theoretical Physics Institute, Department of Physics,
University of Alberta, Edmonton, AB, Canada T6G 2E1, \\
$^{(b)}$ Departamento de Fisica - ICE, Universidade Federal de Juiz de Fora,
Campus da UFJF, CEP: 36036-900, Juiz de Fora, MG, Brazil
}

\abstract{
We discuss some properties of recently proposed models of a ghost-free gravity. For this purpose we
study solutions of linearized gravitational equations in the framework of such a theory. We mainly
focus on the version of the ghost-free theory with the exponential modification
$\exp(\Box/\mu^2)\Box^{-1}$ of the free propagator. The following three problems are discussed: (i)
Gravitational field of a point mass; (ii) Penrose limit of a point source boosted to the speed of
light; and (iii) Spherical gravitational collapse of null fluid.
For the first problem we demonstrate that it can be solved by using the method of heat kernels and
obtain a solution in a spacetime with arbitrary number of dimensions. For the second problem we also
find the corresponding gyraton-type solutions of the ghost-free gravitational equations for any
number of dimensions. For the third problem we obtain solutions for the gravitational field for the
collapse of both "thin" and "thick" spherical null shells. We demonstrate how the ghost-free
modification of the gravitational equations regularize the solutions of the linearized Einstein
equations and smooth out their singularities.
}


\preprint{Alberta Thy 6-15}

\begin{document}

\section{Introduction}

It is widely believed that the quantum gravity will "cure" a decease of the classical general
relativity, its singularities. In particular, in the domain of a spacetime near singularities, where
the curvature becomes large, the Einstein equations should be modified. Such a modification, which,
for example, is dictated by the string theory, should include additional terms in the gravitational
effective action, that are both, higher in curvature and in its derivatives. It was proposed many
different modifications of the Einstein theory of the general relativity, that, in particular,
change its infrared and ultraviolet behavior (see, e.g. a review \cite{Myrzakulov:2013hca}). In this
paper we discuss a special class of the modified gravity theory with higher derivatives. It is
instructive to check at first the effects of such a modification in the linearized version of the
corresponding theory. It is well known that addition of quadratic in the curvature corrections
modifies
the standard Laplace equation for the gravitational potential $\varphi$ in the Newtonian
approximation, which takes, for example, the following form
\be\n{laplap}
(l^2 \lap +1)\lap \varphi=4\pi \rho\, .
\ee
It is easy to show that such a modified equation for a point mass has a decreasing at the infinity
solution, which remains finite at $r=0$. The parameter $l$, which is determined by the coupling
constant for the quadratic in the curvature term in the effective action, provides a UV cut-off in
the regularized solution (see e.g. \cite{Stelle:1977ry,Frolov:1981mz}). A general analysis of the
Newtonian singularities in higher derivative gravity models was recently performed in
\cite{Modesto:2014eta}.

A connected problem is a possibility of mini-black hole production in the gravitational collapse of
a small mass. To study such a problem one may consider it first in the linearized version of the
theory. If a corresponding solution is regular and its curvature for a small mass $M$ is uniformly
small in the whole spacetime, one might conclude that in this regime higher in curvature corrections
are small and can be neglected. In such a case one can expect that for the small enough mass $M$ a
black
hole is not formed. In other words, in such a theory there exists a mass gap for the back hole
formation. Long time ago in the paper \cite{Frolov:1981mz} it was demonstrated that the
gravitational theory with quadratic in the curvature terms in the action possesses this property.

However, in a general case, a propagator in a theory with higher derivatives contains extra poles,
reflecting the existence of the additional to the gravitons degrees of freedom. As a result, the
corresponding theory with higher derivatives usually contains  ghosts
\cite{Stelle:1977ry,Stelle:1976gc,Asorey:1996hz}. In order to 
avoid this problem  a new type of
the modification of the Einstein theory, called ghost-free gravity,
 was proposed
\cite{Biswas:2011ar,Modesto:2011kw,Modesto:2012ys,Biswas:2013cha,Biswas:2013kla}.
 A similar model was proposed earlier in
\cite{Tomboulis:1997gg} (for general discussion see 
\cite{Modesto:2014lga,Shapiro:2015uxa}).

The main idea of this
approach is to consider a theory, which is non-local in derivatives. Suppose that in 
the linearized
version of such a theory the box operator $\Box$ is modified and takes the form 
$a(\Box)\, \Box$,
where  $a(z)$ is an entire function of the complex variable $z$. Then the propagator 
for such a
theory does not have  additional poles, different from the original pole, describing 
propagation of
the gravitons. There exists a variety of entire functions. It is sufficient to use 
the function
$a(z)$ to be an exponent of the polynomial in $z$. The simplest choice is when this 
polynomial is a
linear function, so that the modified box-operator is $\exp(-\Box/\mu^2)\Box$, where 
$\mu$, which
has the dimensionality of the mass, is the UV cut-off parameter.
One may hope that such cut-off
regularizes singularities inside black holes \cite{Modesto:2010uh,Bambi:2013gva} and 
in the big-bang cosmology
\cite{Biswas:2005qr,Khoury:2006fg,Biswas:2010zk,Barvinsky:2012ts,Bambi:2013gva}.
 It should be emphasized that   non-local  operators of a similar form 
and their properties were considered long time ago in the
papers
\cite{efimov1967,Efimov:1972wj,Efimov:1976nu,Efimov:18,Efimov:19,Krasnikov:1987}.

In the present paper we study solutions of the ghost-free gravitational equations in 
the linearized
approximation. In the section~2 we demonstrate that for any number of spacetime 
dimensions $D$ a
solution for the Newtonian potential for a point mass can be  obtained by using the 
heat kernel
method. We describe these solutions and compare them with the solutions of the 
linearized Einstein
equations. We demonstrate that for $D\ge 4$ the gravitational potential is regular 
at the position
of the source, and the parameter $\mu$ is the scale of the UV regularization. We 
also show that for
$D=4$ the obtained result coincides with the one obtained earlier by the Fourier 
method and
presented in the paper \cite{Biswas:2011ar}. In section~3 we boost the solution for 
a static point
mass. By taking the Penrose limit with $v\to c$ of the boosted metric, we obtain a 
solution of the
ghost-free equations for gravitational field of a "photon" moving in $D$-dimensional 
spacetime. We
show that these solutions are similar to the field of non-rotating gyratons
\cite{Frolov:2005in,Frolov:2005zq,FrolovZelnikov:2011}. The main difference  is that 
 a function of
$(D-2)$ transverse variables, which enters the metric of the gyraton and which is a 
solution of the
$(D-2)$ flat Laplace equation, in the ghost-free gravity becomes a solution of the 
$(D-2)$ operator
$a(-\lap)\lap$. As a result, the singularity of the gyraton metric at its origin is 
smooth. In the
next  sections~4--5 we study the problem of the spherical collapse of null 
fluid in the ghost-free
gravity. In section~4 we demonstrate how a solution for the spherical null shell 
collapse can be
obtained as a superposition of a spherical distribution of gyratons, that pass 
through the fixed
point and are the generators of the null shell. The method can be used in a 
spacetime with an
arbitrary number of dimensions. In the present paper we illustrate it for a special 
case of the four
dimensional spacetime. We use this approach to obtain a solution for the 
gravitational
field of a thin null shell in the ghost-free theory. We show that a solution, 
obtained in the
developed gyraton-based  approach, correctly reproduces the known solution for the 
collapsing null
shell in the linearized Einstein gravity. We obtain also a solution for the 
ghost-free gravity and
compare it with the solution for the Einstein gravity. In particular, we calculate 
the
Kretschmann curvature invariant for the solution and demonstrate that its 
singularity is
smoothened. However, it
remains divergent at the origin $r=0$. Finally, we study the collapse of a spherical 
thick null
shell and obtain its gravitational field by taking a specially chosen superposition 
of thin
null shells (section~5). We demonstrate that the ghost-free gravitational 
field is everywhere regular in this
model, and for small mass $M$ the apparent horizon (and hence a  black hole) does 
not form. This
means that in the ghost-free gravity there is a mass gap for the mini black hole 
formation in the
gravitational collapse, and, in this sense, its properties are somehow similar to 
the properties of
the theory with quadratic in curvature corrections, discussed in 
\cite{Frolov:1981mz}. Section~6
contains summary and discussions of the obtained results.  Appendices contain 
details of the derivation of the thin null shell metric from the gyraton solutions.
In this paper we use the sign convention adopted in the book \cite{Misner:1974qy}.

\section{Non-local Newtonian gravity}

\subsection{Newtonian potential in the ghost-free gravity}

In this paper we consider solutions of the linearized ghost-free gravity equations.
We start by analyzing static solutions for a point mass in this approximation. Such 
a solution gives
a modified Newtonian potential. The gravitational field of a point source in four 
dimensions was
obtained earlier in \cite{Biswas:2011ar,Modesto:2010uh}.
The authors of these papers used  the Fourier method for
this purpose. We demonstrate that this result can be obtained much easier by using 
the method of the
heat kernels. This method, practically without any changes, allows one to solve a 
similar problem in
the higher dimensional case.
We consider a flat spacetime and denote by $D$ the number of its dimensions. It is 
often convenient
to use a number $n$, connected with $D$ as follows $D=n+3$. The static metric in the 
weak-field
approximation in the standard $D$ dimensional gravity  can be written in the form 
(see e.g.
\cite{FrolovZelnikov:2011})
\be
ds^2=-(1+2\varphi)dt^2 +\left(1-{2\over n}\varphi\right) d\ell^2\hhh
d\ell^2=\delta_{ij}dx^i dx^j\hhh i,j = 1,\ldots,D-1
\, ,
\ee
where the Newtonian potential $\varphi$ satisfies the equation
\be\n{ggp}
\lap \varphi =4\pi G \rho\hh G={2n\over n+1}G^{D}\, .
\ee
The operator $\lap$ is a standard  $(D-1)$-dimensional Laplace operator, which 
in the flat (Cartesian) coordinates is of
the form
\be
\lap(\ldots)=\delta^{ij}(\ldots)_{,ij}\, .
\ee
In the four-dimensional spacetime $n=1$, so that $G=G^4$ and (\ref{ggp}) takes the 
standard form of
the Poisson equation for the Newtonian gravitational potential.
It should be emphasized that there exists an ambiguity in the choice of the form of 
the higher
dimensional coupling constant. We denote this constant by $G^D$ and fixed it by the 
requirement that
the form of the Einstein equation is the same in any number of dimensions
\be
S[g]={1\over 16\pi G^D}\int d^Dx \sqrt{-g}R\, .
\ee
We introduce another constant $G$, related with $G^D$ in such a way, that the form 
of the equation
for the Newtonian potential $\varphi$ is the same for any $D$. For a point mass $m$
\be\n{sour}
\rho=m\delta^{n+2}(x)\, ,
\ee
and the potential $\varphi$ is (see e.g. \cite{FrolovZelnikov:2011})
\be
\varphi=-{2\Gamma(1+n/2)\over \pi^{n/2} n}{Gm\over r^n}\, .
\ee
In the four-dimensional case this expression takes the standard form $\varphi=-Gm/r$.

Following \cite{Biswas:2011ar} we write the modified ghost-free equation for the 
Newtonian
gravitational potential $\tilde{\varphi}$, created by a point source of the mass 
$m$, in the form
\be\n{eqq}
\hat{F} \tilde{\varphi}=4\pi G \rho\hh \hat{F}=a(\lap)\lap\, .
\ee
We shall use the following form of the non-local operator $a(\lap)$, proposed in
\cite{Biswas:2011ar}
\be\n{aaa}
a(\lap)=e^{-\lap/\mu^2}\, .
\ee
The parameter $\mu$ specifies the characteristic energy scale, where the adopted 
theory modification becomes important. In what follows we shall also use another 
parameter, $s$, related with $\mu$ as follows
\be\n{smu}
s=\mu^{-2}\, .
\ee
This is convenient, since some of the relations will contain derivatives and 
integrals over $s$, which do not look "elegant" in terms of $\mu$. On the other 
hand, $\mu$ has more direct "physical meaning" as the mass (energy) of the UV 
cut-off of the modified gravity. After required calculations are performed one can 
simply substitute the parameter $s$ in terms of $\mu$ by using the relation 
(\ref{smu}).

\subsection{Heat kernel approach}

Let us discuss now equation (\ref{eqq}) and its solution. We shall use the following 
standard notions. For any operator $\hat{B}$ we denote by $B(x,x')$ its value in the 
$x$-representation
\be
B(x,x')=\langle x|\hat{B}|x' \rangle\, .
\ee
In these notations $\delta$-function is
\be
\delta(x,x')=\langle x|\hat{I}|x' \rangle\, ,
\ee
where $\hat{I}$ is a unit operator. Using these notations one can write the equation 
(\ref{eqq}) as follows
\be
\int dx'\, F(x,x') \tilde{\varphi}(x')=4\pi G \rho(x)\, .
\ee
We denote by $\hat{A}$ an inverse to $\hat{F}$ operator
\be\n{AFF}
\hat{A}=\hat{F}^{-1}={1\over \lap} e^{s\lap}=e^{s\lap}{1\over \lap}\, .
\ee
We denote
\be\n{oper}
\hat{D}=-{1\over \lap}\hh \hat{K}(s)=e^{s\lap}\, ,
\ee
so that relation (\ref{AFF}) takes the form
\be
\hat{A}=-\hat{D}\hat{K}(s)=-\hat{K}(s)\hat{D}\, .
\ee
In the $x$-representation $D(x,x')$ is nothing but a usual Green function of the 
Laplace operator and it obeys the equation
\be
\lap \hat{D}=-\hat{I}\, .
\ee
The other operator, $\hat{K}(s)$, is also well known. In $x$-representation it is 
just a standard heat kernel $K_n(x,x'|s)$ of the Laplace operator. It has the 
following form
\be\n{KK}
K_n(x,x'|s)\equiv {1\over (4\pi s)^{1+n/2}}e^{-\lambda(x,x')/(4s)}\, .
\ee
This formula requires some explanations. We remind that we denoted by $D=n+3$ the 
total number of spacetime dimensions. Since the gravitational field of a static 
source is static, the gravitational potential depends only on $D-1=n+2$ spatial 
coordinates. Correspondingly, the heat kernel (\ref{KK}) is a function of 2 spatial 
points, $x$ and $x'$ and $\lambda(x,x')$ is the square of the spatial distance 
between these points
\be
\lambda(x,x')=|x-x'|^2\, .
\ee
The degree $1+n/2$ of expression in the denominator reflects the fact that we are 
working in a space with $D-1=n+2$ dimensions.

It is easy to check that the operator $\hat{A}$ can be written in terms of 
$\hat{K}(s)$ as follows
\be\n{AAK}
\hat{A}_n=-\int_s^{\infty} ds\, \hat{K}_n(s)\, .
\ee
Relations (\ref{KK}) and (\ref{AAK}) allow one to obtain a solution of the equation 
(\ref{eqq})-(\ref{aaa}). For a point mass $m$ at point $x'=0$  one has
\be
\tilde{\varphi}(x)=4\pi m A_n(x,0)=-4\pi G m \int_s^{\infty} ds\, \hat{K}_n(x,0|s)\, 
.
\ee

Before presenting explicit form of the solutions for the Newtonian potential 
$\tilde{\varphi}$ in the ghost-free gravity, let us make the following remark. The 
equations (\ref{eqq})--(\ref{aaa}) can be identically rewritten in the form
\be
\lap \tilde{\varphi}=4\pi G \tilde{\rho}\hh
\tilde{\rho}=\hat{K}_n(s) \rho\, .
\ee
In other words, the Newtonian potential $\tilde{\varphi}$ in the ghost-free gravity 
is a solution of
the standard Poisson equation for the source $\tilde{\rho}$, which is obtained by 
smearing the
original distribution $\rho(x)$. For the point mass (\ref{sour}) the smeared source 
is
\be
\tilde{\rho}(x)=m K_n(x,0|s)\, .
\ee
In this approach one may say that the ghost-free gravity regularizes the potential 
$\tilde{\varphi}$ by smearing a source, which generates it. Of course, the results 
obtained by the source smearing and by applying $\hat{A}$ operator to "un-smeared" 
source $\rho$ are the same. Let us present the results.

{\bf In flat three-dimensional space} $(D=4, n=1)$ in Cartesian coordinates
\ba
&&K_1(x,x'|s)={1\over (4\pi s)^{3/2}}e^{-{\lambda\over 4s}}\, ,\\
&&A_1(x,x')=-{1\over 4\pi \sqrt{\lambda}}\mbox{erf}\left({\sqrt{\lambda}\over 
2\sqrt{s}}\right)\, ,\\
&&\tilde{\varphi}=-{Gm \over r}\mbox{erf}\left( {\mu r\over 2}\right)\, . 
\n{fff4}
\ea
Here $\mbox{erf}(x)$ is the Gauss error function, which is defined as follows
\be
\mbox{erf}(x)={2\over \sqrt{\pi}}\int_0^x\, e^{-t^2} dt\, .
\ee
Formula (\ref{fff4}) correctly reproduces the expression for the gravitational 
potential in the linearized ghost-free gravity, obtained earlier 
\cite{Biswas:2011ar,Modesto:2010uh}.

{\bf In flat four-dimensional space} $(D=5, n=2)$ in Cartesian coordinates
\ba
&&K_2(x,x'|s)={1\over (4\pi s)^{2}}e^{-{\lambda\over 4s}}\, ,\\
&&A_2(x,x')=-{1\over 4\pi^2 \lambda}\left[1-e^{- \lambda/(4s)}\right]\, ,\\
&&\tilde{\varphi}=-{Gm \over \pi r^2}\left( 1-\exp(-\mu^2 r^2/4)\right)\, .
\ea

There exist simple relations between $K_n(x,x'|s)$ and $A_n(x,x')$ in spacetimes 
with different number of dimensions
\be
K_{n+2}(x,x'|s)={1\over\pi}{\partial \over \partial \lambda}K_n(x,x'|s)\hhh
A_{n+2}={1\over\pi}{\partial \over \partial \lambda}\,A_n(x,x')\, .
\ee
Using these relations one can  obtain a general expression for $A_n(x,x')$
\be
A_n(x,x')=-{\gamma\left({n/2},{\lambda/(4s)}\right)\over  
4\pi^{(n+1)/2}\lambda^{n/2}}\hh n\ge 2\, .
\ee
This relation contains a so called  lower incomplete gamma function $\gamma(a,x)$, 
which is is defined as
\be
\gamma(a,x)=\int_0^x \ t^{a-1} e^{-t} dt=
x^a \Gamma(a) e^{-x} \sum_{k=0}^{\infty}{ x^k\over \Gamma(a+k+1)}\, .
\ee

Thus in the ghost-free gravity the Newtonian field of a point mass $m$ is
\be\n{ffff}
\tilde{\varphi}(r)=- G m {\gamma\left({n/2},{r^2/(4s)}\right)\over  \pi^{(n-1)/2} 
r^{n}}\, .
\ee
At small $x$ one has $\gamma(a,x)\approx x^a/a$. Hence
\be\n{Ph0}
\tilde{\varphi}(r=0)=-{2Gm\over n \pi^{(n-1)/2} (4s)^{n/2}}\, .
\ee
This relation implies that the Newtonian potential of a point mass in the ghost-free 
gravity in any number of dimensions $D\ge 4$ is finite at the origin. In other 
words, this potential is properly regularized.

\section{Non-spinning gyratons in the ghost-free gravity }

We obtain now the gravitational field of an ultra-relativistic particle in the 
framework of the
ghost-free gravity. Instead of solving the modified gravitational equations for a 
source moving with
the speed of light we use the following procedure which is well known in the 
standard General
Relativity. We first make the Lorentz transformation of the  static solution for a 
point mass $m$
and obtain the metric for the  object moving with velocity $\beta$. After this we 
take a so called
Penrose limit of the metric of the moving body. Namely, we take the limit $\beta\to 
1$,
while keeping the
energy of the object $\gamma m$ fixed. As a result one obtains an object in 
$D$-dimensional
spacetime which was called a gyraton 
\cite{Frolov:2005in,Frolov:2005zq,FrolovZelnikov:2011}. In a
general case, when a static source has an angular momentum, the corresponding 
gyraton is spinning.
We
restrict ourselves by the case of non-spinning gyratons. A new element in our 
derivation is
performing the described procedure not within the General Relativity, but in the 
ghost-free theory
of gravity.

To perform the calculations it is convenient to use the following notations for the 
standard
Cartesian coordinates $x^{\mu}=(\bar{t},y,\BM{\zeta}_{\inds{\perp}})$. Here 
$\bar{t}$ is just the
time in the frame, where the source is at rest, $y$ is the coordinate in the 
direction of motion of
the source, and $\BM{\zeta}_{\inds{\perp}}=(\zeta^2,\ldots,\zeta^{D-2})$ are 
the coordinates in
$(D-2)$ dimensional plane orthogonal to the direction of the motion. We shall call 
the latter the
{\em transverse} coordinates. The Newtonian potential for a point source of mass $m$ 
in the
ghost-free gravity, obtained in the previous section, takes the following form in 
these coordinates
\ba
ds^2&=&ds_0^2+dh^2\, ,\\
ds_0^2&=& -d\bar{t}^2+dy^2+ d\zeta_{\inds{\perp}}^2\hh
d\zeta_{\inds{\perp}}^2=\sum_{2}^{n+1} (d\zeta_i)^2\, ,\\
dh^2&=&-2\tilde{\varphi}\left[ d\bar{t}^2+{1\over 
n}(dy^2+d\zeta_{\inds{\perp}}^2)\right]\, ,
\ea
where $\tilde{\varphi}$ is defined by (\ref{ffff}).

To obtain a metric for a  source moving with the velocity $\beta$ (along the 
$y$-axis in the
positive direction) we make the following Lorentz transformation \be\begin{split}
y&=\gamma (\xi-\beta t)={\gamma \over 2} [(1-\beta) v-(1+\beta)u]\, ,\\
\bar{t}&=\gamma (t-\beta \xi)={\gamma \over 2} [(1-\beta) v+(1+\beta)u]\, .
\end{split}
\ee
Here $(t,\xi,\BM{\zeta}_{\inds{\perp}})$ are Cartesian coordinates in the new 
inertial frame, where
the source is moving. We also denoted \be
\gamma=(1-\beta^2)^{-1/2}\hhh u=t-\xi\hhh v=t+\xi\, .
\ee
The flat metric $ds_0^2$ in the new coordinates 
$x^{\mu}=(u,v,\BM{\zeta}_{\inds{\perp}})$ is
\be
ds_0^2=-du\, dv+ d\zeta_{\inds{\perp}}^2\, .
\ee
The form of this metric remains the same in the limit $\beta\to 1$, while in this 
limit
\ba
&&y\sim -\gamma u\hh \bar{t}\sim \gamma u\hh
\lambda(x,0)\sim \gamma^2 u^2+\zeta_{\inds{\perp}}^2\, ,\\
&&d\bar{t}^2+{1\over n}(dy^2+d\zeta_{\inds{\perp}}^2)\sim
{n+1\over n}\gamma^2 du^2+\ldots\, .
\ea
Here $\ldots$ denote sub-leading in $\gamma$ terms. As a result, the metric 
perturbation in this
limit takes the form \be
dh^2=\Phi du^2\hh
\Phi=-{2(n+1)\over n}\lim_{\gamma\to \infty} (\gamma^2 \tilde{\varphi})\, .
\ee
When taking this limit, we assume, as usual, that the energy of the object, $\gamma 
m$ is fixed (the
Penrose limit), and we denote \be
M=G\gamma m\, .
\ee
We also use the following relation
\be
\lim_{\gamma\to \infty} \gamma e^{-{\gamma^2 u^2\over 4s}}=\sqrt{4\pi s} \delta(u)\, 
.
\ee
Combining these results, we finally obtain the following expression for the function 
$\Phi$
\be
\Phi=-8\pi {n+1\over n} M \delta(u) A_{n-1}(\zeta_{\inds{\perp}})\, ,
\ee
where
\be\n{ann}
A_{n-1}(\zeta_{\inds{\perp}})=-\int_s^{\infty} {ds\over (4\pi s)^{ {n+1\over 2}}}\,
e^{-\zeta_{\inds{\perp}}^2/(4s)}\, . \ee

Let us notice, that $A_{n-1}(\zeta_{\inds{\perp}})$ is nothing, but a solution of 
the ghost-free
gravity equations for a point source, reduced to $D-2$ dimensional transverse plane 
with coordinates
$\BM{\zeta}_{\inds{\perp}}$. For large $\mu$ it reduces to a solution of the $D-2$ 
dimensional
Poisson equations. This means, that in this limit the obtained solution of the 
ghost-free gravity
equations reduces to the standard metric of non-rotating gyratons
\cite{Frolov:2005in,Frolov:2005zq,FrolovZelnikov:2011}.

In what follows we restrict ourselves by considering a four dimensional spacetime. 
For this case the
integral, which enters the definition of $A_0$, has an infrared divergence at large 
$s$. Let us
discuss
this case in more detail. Let us notice that \be
A_0=-{1\over 4\pi}\int_s^{\eta^2} {ds\over s}e^{-\zeta_{\inds{\perp}}^2/(4s)}=
{1\over 4\pi} \left[  \mbox{Ei} (1,\zeta_{\inds{\perp}}^2/(4s)+\mbox{Ei}
(1,\zeta_{\inds{\perp}}^2/(4\eta^2)\right]\, . \ee
The parameter $\eta$, which has the dimensionality of the length, is an infra-red 
cut-off parameter.
Here $\mbox{Ei}(a,z)$ is the {\em exponential integral} defined as \be
\mbox{Ei}(a,z)=\int_{1}^{\infty} dx\, x^{-a}\, e^{-x z}\, .
\ee
The function $\mbox{Ei}(1,z)$ for small $z$ has the following expansion
\be\n{laz}
\mbox{Ei}(1,z)=-\ln(z) -\gamma +\ldots\, .
\ee
Here $\gamma=0.5772156649$ is the Euler constant, and $\ldots$ denote the terms 
vanishing in the
$z=0$ limit. Assuming that $\eta^2$ is large and using (\ref{laz}) one can write
\be
A_0={1\over 4\pi}\left[ \ln(\zeta_{\inds{\perp}}^2/\eta^2) +\gamma+ \mbox{Ei}
(1,\zeta_{\inds{\perp}}^2/(4s)\right]\, . \ee
The corresponding expression for the metric is
\ba
&&ds^2=-du dv +d\zeta_{\inds{\perp}}^2+dh^2\hh
dh^2=\Phi du^2\, ,\n{hhhh}\\
&&\Phi=-4M \delta(u) F(\zeta_{\inds{\perp}}^2)\, ,\\
&&F(\zeta_{\inds{\perp}}^2)=\ln(\zeta_{\inds{\perp}}^2/\eta^2) +\gamma+ \mbox{Ei}
(1,\zeta_{\inds{\perp}}^2/(4s))\, .\n{soll} \ea
In the limit $\mu\to \infty$, when the ghost-free gravity reduces to the Einstein 
theory,
\be
F(\zeta_{\inds{\perp}}^2)=\ln(\zeta_{\inds{\perp}}^2/\eta^2) +\gamma\, .
\ee
This result reproduces the well known Aichelburg-Sexl solution 
\cite{Aichelburg:1970dh} (see also
\cite{Frolov:1998wf}).

Let us notice that the solution (\ref{soll}) contains an arbitrary parameter, the 
infrared cut-off
parameter $\eta$. However, the change of this parameter can be easily absorbed into 
the redefinition
of the advanced time $v$. Hence, this ambiguity just reflects freedom in the gauge 
choice. In order
to demonstrate this, let us consider a metric of the form \be
ds^2=-du dv +d\zeta_{\inds{\perp}}^2+f(u) du^2\, .
\ee
Let us define a new coordinate
\be
\bar{v}=v-\int du f(u)\, .
\ee
Then one gets
\be
ds^2=-du d\bar{v} +d\zeta_{\inds{\perp}}^2\, .
\ee
This confirms our above conclusion, that the ambiguity in the choice of the cut-off 
parameter
$\eta$
can always been absorbed in the change of the coordinates. In what follows we use 
this option and
simply put  $\eta^2=4s$. For this choice \be
F(z)=\ln(z) +\gamma+ \mbox{Ei} (1,z)\hh
z=\zeta_{\inds{\perp}}^2/(4s)\, .\n{Fsoll}
\ee
For this choice the expansion of the function $F(z)$ for small $z$ takes a very 
simple form
\be\n{Fexp}
F(z)=z-{1\over 4}z^2 +O(z^3)\, .
\ee

\section{Null shell collapse}

\subsection{Null shell as a superposition of null gyratons}

We use now the above described gyraton metric in order to study the gravitational 
collapse in the
ghost-free gravity. Namely, we consider a collapse of a spherical null shell. As 
earlier we use
linearized gravitational equations. For simplicity, we restrict ourselves by 4D 
case. Instead of
solving the corresponding equations we shall use the following trick. Let us notice 
that a sum of
solutions of the linearized theory is again a solution. In the linear approximation 
the solution
has
the form
\be\n{met}
ds^2=ds_0^2+dh^2\, .
\ee
Here $ds_0^2$ is the flat metric and $dh^2$ is a perturbation.

Let us consider a set of gyratons passing through a chosen point $O$ of the 
spacetime.
We use  the Cartesian coordinates $(t,X,Y,Z)$ in the the Minkowski spacetime, and 
identify a point
$O$ with the origin of the coordinate system $O=(0,0,0,0)$. We denote by
$(\BM{e}_{\inds{X}},\BM{e}_{\inds{Y}},\BM{e}_{\inds{Z}})$ unit vectors in the 
directions of the axes
$X$, $Y$, and $Z$, respectively, and by $\BM{n}$ a unit vector in 3D space in the 
direction of the
motion of a fixed gyraton. One has \be\n{nn}
\BM{n}=\sin\alpha \, (\cos\beta ~\BM{e}_{\inds{X}} +\sin\beta 
~\BM{e}_{\inds{Y}})+\cos\alpha ~
\BM{e}_{\inds{Z}}\, .
\ee
Here $(\alpha,\beta)$ are standard coordinates on a unit 2D sphere. Let $a$ be an 
index enumerating
the gyratons and the metric perturbation created by such a gyraton is $dh_a^2$. Then 
the
perturbation  $\langle dh^2\rangle$ created by a set of the gyratons has the 
form \be\n{sum}
\langle dh^2\rangle =\sum_a dh_a^2\, .
\ee
We assume that all gyratons have the same energy. We also take a continuous limit of 
the
discrete distribution of the gyratons and assume that such a distribution is 
spherically symmetric.
Thus, we write \be\n{ave}
\langle dh^2\rangle={1\over 4\pi}\int_0^{\pi} d\alpha\, \sin\alpha \int_0^{2\pi} 
d\beta \ \
dh^2_{(\alpha,\beta)}\, . \ee
An extra factor $(4\pi)^{-1}$ reflects that we use averaging over a unit sphere.

As a result of the averaging, the source of the metric $dh^2$ is a thin null shell 
located at the
null cones $\Gamma_{\pm}$ \be
t=\pm  \sqrt{X^2+Y^2+Z^2}\, .
\ee
For the sign minus, $\Gamma_-$ is a null cone with apex at $O$, which describes a 
collapsing
spherical null shell. For the sign plus, the null cone $\Gamma_+$ describes an 
expanding null shell.
Our starting point is gyraton metric (\ref{hhhh}), which we rewrite in the form
\be
ds_0^2=-dt^2+d\xi^2+d\zeta_{\inds{\perp}}^2\hh
dh^2=\Phi (dt-d\xi)^2\hh \Phi=-4M F(\zeta_{\inds{\perp}}^2)\delta(t-\xi)\, .
\ee
Here
\be
F=\left\{ \begin{array}{ll}
\displaystyle{\ln(\zeta_{\inds{\perp}}^2/\eta^2)\, ,} & \mbox{for the Einstein 
theory}\, ;\\
\displaystyle{\mbox{Ei}(1,\zeta_{\inds{\perp}}^2/4s)+\gamma 
+\ln(\zeta_{\inds{\perp}}^2/4s)\, ,}
& \mbox{for the ghost-free gravity} \, .
\end{array}\right.
\ee
Let us notice that this is a metric of a gyraton without spin. Namely these 
gyratons will be considered
in this paper. We also found convenient to use the following terminology. We call a 
worldline of a
gyraton a null string. Its projection to the 3D space, the gyraton trajectory, is an 
oriented
straight line. A unit orientation vector $\BM{n}$ along this line determines a 
direction of the
motion. Quantity $\xi$ is the coordinate along the trajectory in the direction of 
motion of the
"photon", and $\zeta_{\inds{\perp}}=(\zeta_1,\zeta_2)$ are Cartesian coordinates in 
the 2D plane
orthogonal to this direction. We study a spherically symmetric distribution of the 
gyratons, which
has the property that the null strings, representing them, intersect at a single 
spacetime point
$O$. We also choose the parameter $\xi$ along each of the strings to vanish at this 
point $O$.
Consider a point $P=(X,Y,Z)$ of the 3D space. There exist exactly two gyraton 
trajectories, passing
through this point. Let $(\alpha_+,\beta_+)$ be angles of a unit vector $\BM{n}_+$ 
(see (\ref{nn}))
along the line, connecting the origin with the point $P$. The parameter $\xi_+$ at 
$P$ for such a
trajectory is positive. The direction vector  for the second trajectory is 
$\BM{n}_-=-\BM{n}_+$ and
its angles are $(\alpha_-=\pi-\alpha_+,\beta_-=\pi+\beta_+)$, while the 
corresponding coordinate
$\xi_-=-\xi_+$ is negative.

It is convenient to perform the calculations of $\langle dh^2\rangle$  in two steps. 
First we
introduce the following objects \ba
&&T^{(\alpha,\beta)}(y_{\inds{\perp}})=\delta^2(y_{\inds{\perp}}-\zeta_{\inds{\perp}}
)\delta(t-\xi)
(dt-d\xi)^2\, ,\\ &&\langle T(y_{\inds{\perp}})\rangle={1\over 4\pi}\int_0^{\pi} 
d\alpha \sin\alpha
\int_0^{2\pi} d\beta \n{Ty}\ \ T^{(\alpha,\beta)}(y_{\inds{\perp}})\, .\n{VV}
\ea
For $y_{\inds{\perp}}=0$
\be
T^{(\alpha,\beta)}=T_{\mu\nu}^{(\alpha,\beta)}dx^{\mu}dx^{\nu}
\ee
is the stress-energy tensor of a gyraton moving in the direction $\BM{n}$ (see 
(\ref{nn})). Similarly, for $y_{\inds{\perp}}=0$
\be\n{tns}
M\langle T\rangle=T_{\mu\nu}dx^{\mu}dx^{\nu}\, ,
 \ee
where $T_{\mu\nu}$ is the stress-energy tensor of a spherical null shell, 
constructed from gyraton
null strings. Next, we use $\langle T(y_{\inds{\perp}})\rangle$ to find the metric 
perturbation
$\langle dh^2\rangle$ for the thin null shell of mass $M$ \be\n{rep}
\langle dh^2\rangle =M \int dy_{\inds{\perp}} F(y_{\inds{\perp}}^2)\langle 
T(y_{\inds{\perp}})\rangle\, .
\ee

It should be emphasized that the quantities $dh^2_{(\alpha,\beta)}$ and
$T^{(\alpha,\beta)}(y_{\inds{\perp}})$, which enter (\ref{ave}) and (\ref{VV}), must 
be first
written in the coordinate system, that does not depend on the particular value of 
the parameters
$(\alpha,\beta)$. We use the Cartesian coordinates $(X,Y,Z)$ for this purpose.
Thus we need first to
establish relations between the gyraton associated coordinates 
$(\xi,\zeta_{\inds{\perp}})$ and the
Cartesian coordinates $(X,Y,Z)$. This problem is solved in the appendix~A. After 
this we need to calculate the integral over the sphere, which enters relation 
(\ref{VV}) for the average value
$\langle T(y_{\inds{\perp}})\rangle$. The details of these calculations are 
collected in appendix~B.

\subsection{Stress-energy tensor}

The relation (\ref{aver}) can be used to find the stress-energy tensor of the null 
shell
constructed from null strings representing a set of gyratons. Using (\ref{Ty}) and 
(\ref{aver})  one
gets 
\be
\langle T(y_{\inds{\perp}})\rangle={1\over 4\pi |Q\xi|}\left[
\sin\alpha_+\, \delta(t-\xi_+)\, (dt-d\xi_+)^2+
\sin\alpha_- \, \delta(t-\xi_-)\,  (dt-d\xi_-)^2\right]_{y_{\inds{\perp}}}\, .
\ee
Taking the limit $y_{\inds{\perp}}\to 0$ in this relation and using relations 
(\ref{tns}) and
(\ref{scay}) one obtains \be
M\langle T\rangle=T_{\mu\nu}dx^{\mu}dx^{\nu}={M\over 4\pi r^2}
[ \delta(u) du^2+\delta(v) dv^2]\, ,
\ee
where $u=t-r$, $v=t+r$. This relation correctly reproduces the expected expression 
for the
stress-energy tensor of the null shell. It is a superposition of the stress-energy 
tensors of
contracting and expanding spherical null shells of mass $M$.
One can easily solve the linearized Einstein equations for such a null-shell 
problem. The solution
is well known and simple. Inside both the collapsing and expanding shells the 
spacetime is flat,
while outside them the metric is a linearized version of the Schwarzschild metric 
\be\n{sch}
\langle dh^2\rangle={2M\over r}(dt^2+dr^2)\, .
\ee
Our next goal is to reproduce this result by using the representation (\ref{rep}). 
This will provide
us with a useful test of the validity of our approach.
\subsection{Averaged metric}

We have all the required expressions to perform the calculations. However, one can 
greatly simplify
the problem using the following observation. Since the distribution of the null 
strings representing
gyratons is spherically symmetric, the corresponding averaged metric $\langle 
dh^2\rangle$ must also
have this property. Hence, it can be written in the form \be\n{ssm}
\langle dh^2\rangle=h_{tt} dt^2+2 h_{tr} dt dr +h_{rr} dr^2+ H d\omega^2\, .
\ee
Here the metric coefficients $h_{tt}$, $h_{tr}$, $h_{rr}$ and $H$ are functions of 
$t$ and $r$, and
$d\omega^2$ is the metric on a unit sphere. The spherical coordinates 
$(r,\theta,\phi)$ are related
with the Cartesian coordinates $(X,Y,Z)$ as follows \be
X=r\sin\theta\cos\phi\hh Y=r\sin\theta\sin\phi\hh Z=r\cos\theta\, .
\ee
In order to find the metric (\ref{ssm}) it is sufficient to obtain its value near a 
single spatial
point. We choose such a point as follows: $P_{\inds{X}}=(X=r,0,0)$, where $X>0$. For 
this choice the
metric (\ref{ssm}) reduces to 
\be\n{XX}
\langle dh^2\rangle=h_{tt} dt^2+2 h_{tr} dt dX +h_{rr} dX^2+ {H\over r^2}\,  
\left(dY^2+dZ^2\right)\, .
\ee
We perform now calculations of $\langle dh^2\rangle$ at $P_X$ and by comparing it 
with (\ref{XX}) we
find the expression for the metric perturbation (\ref{ssm}).
One has near $P_{\inds{X}}$
\ba
&&Q_{\pm}=\pm\sqrt{X^2-y_p^2}\hh \xi_{\pm}=\pm \sqrt{X^2-y_{\inds{\perp}}^2}\, ,\\
&&\sin\alpha_{\pm}={\xi_{\pm}\over Q_{\pm}}\hh \cos\alpha_{\pm}=-{y_k\over 
Q_{\pm}}\, ,\\
&& \sin\beta_{\pm}={y_p\over X}\hh \cos\beta_{\pm}={Q_{\pm}\over X}\, ,\\
&& d\xi_{\pm}={\xi_{\pm}\over X} dX+{1\over Q_{\pm}}\left(\xi_{\pm}\,{y_p\over X}\, 
dY-y_k\, dZ\right)\, .
\ea

In order to obtain the metric perturbation $\langle dh^2\rangle$ one must to 
calculate the integral
over $(y_p,y_k)$ in (\ref{rep}). We introduce polar coordinates in the 
$(y_p,y_k)$-plane: \be
y_p=\rho\cos\psi\hh y_k=\rho\sin\psi\, .
\ee
We also write
\ba
&&\langle T(y_{\inds{\perp}})\rangle =T^++T^-\, ,\\
&& T^{\pm}={1\over 4\pi Q^2} \,  \delta(t-\xi_{\pm}) \, (dt-d\xi_{\pm})^2\, .
\ea
Because of the presence of $\delta$-function the term $T^+$ is non-zero only for 
$t\ge 0$, while the
other term $T^-$ is non-zero for $t\le 0$. We write expression (\ref{rep}) in the 
form \ba\n{intt}
&&\langle dh^2\rangle= -4M\int_0^{\infty} d\rho \rho F(\rho^2) ({\cal T}^+ + {\cal 
T}^-)\, ,\\
&&{\cal T}^{\pm}=\int_0^{2\pi} d\psi T^{\pm}\n{psi}\, .
\ea
One has
\ba
&&(dt-d\xi_{\pm})^2=A_{\pm}+{B_{\pm}\over Q^2}+\ldots\hh Q^2=X^2-\rho^2\cos^2\psi\, 
,\\
&&A_{\pm}=(dt-{\xi_{\pm}\over X} dX)^2\hh
B_{\pm}=(a_{\pm}\cos\psi +b\sin\psi)^2\, ,\\
&&a_{\pm}={\xi_{\pm}\rho dY\over X}\hh
b=-\rho dZ\, ,
\ea
and dots indicate terms linear in $y_p$ and $y_k$, which do not contribute to the 
integral over
$\psi$ (\ref{psi}). Thus one has \be\n{ttt}
{\cal T}^{\pm}={\delta(t-\xi_{\pm})\over 4\pi}\, (A_{\pm}\,  K_1+K_2)\hhh 
K_1=\int_0^{2\pi}
{d\psi\over Q^2}\hhh K_2=\int_0^{2\pi} {d\psi\, B_{\pm}\over Q^4}\, . \ee
Integrals $K_1$ and $K_2$ can be easily taken with the following result
\ba
K_1&=&{2\pi\over X \sqrt{X^2-\rho^2}}\, ,\\
K_2&=&{\pi [(a_{\pm}^2+b^2) X^2-b^2\rho^2]\over X^3(X^2-\rho^2)^{3/2}}={\pi 
\rho^2\over X^3
\sqrt{X^2-\rho^2}}(dY^2+dZ^2)\, . \ea
$\delta$-function in (\ref{ttt}) can be written as
\be
\delta(t\mp \sqrt{X^2-\rho^2})={|t|\over \rho}\delta(\rho-\sqrt{X^2-t^2})\, .
\ee
This relation also implies that
\be
t=\pm \sqrt{X^2-\rho^2}\, .
\ee
After integration in (\ref{intt}) the terms ${\cal T}^{\pm}$ give similar 
contributions, so that
finally one obtains the following result \be\n{dhX}
\langle dh^2\rangle={-2M F(X^2-t^2)\over X}\left[ \left(dt-{t\over X}dX\right)^2 
+{X^2-t^2\over
2X^2}(dY^2+dZ^2)\right]\, . \ee
Comparing (\ref{dhX}) with (\ref{XX}) we finally get
\be\n{dhr}
\langle dh^2\rangle={-2M F(r^2-t^2)\over r}\left[ \left(dt-{t\over r}dr\right)^2 
+{r^2-t^2\over
2}d\omega^2\right]\, . \ee
This expression is valid everywhere in the domain $r \ge |t|$ for both positive and 
negative time $t$.

Using GRTensor program one can check that the Ricci tensor for this perturbation of 
the  flat metric
in the linear in $M$ approximation vanishes everywhere outside the null shell. This 
provides one
with a good test of the correctness of the performed calculations.

\subsection{A case of linearized Einstein equations}

For the linearized Einstein equations
\be\n{fln}
F(r^2-t^2)=\ln\left( {r^2-t^2\over \eta^2}\right)\, .
\ee
The corresponding expression (\ref{dhr}) looks quite different from the expected 
answer
(\ref{sch}).
However one can use the gauge freedom
\be
h_{\mu\nu}\to h_{\mu\nu}-V_{(\mu;\nu)}\, ,
\ee
to transform (\ref{dhr}) into the expected form. It is sufficient to choose
\ba
&&V_{\mu}=(V_t,V_r,0,0)\, ,\\
&&V_t=-{2Mt\over r}(1+t/r)\ln\left( {r^2-t^2\over\eta^2}\right)+2M\ln{r-t\over 
r+t}+{2Mt\over r}\, ,\\
&&V_r=-{M\over r^2}(r^2-t^2)\ln\left( {r^2-t^2\over \eta^2}\right)\, .
\ea
After this gauge transformation one gets
\be
\langle dh^2\rangle={2M\over r}(dt^2+dr^2)\, .
\ee
Let us notice that the infrared cut-off $\eta^2$ in (\ref{fln}) does not enter the 
final result and
the change of this parameter is simply absorbed into the redefinition of the gauge 
field $V_{\mu}$.
Let us mention also that the Kretschmann invariant
\be
{\cal R}^2=R_{\mu\nu\alpha\beta}R^{\mu\nu\alpha\beta}
\ee
in its lowest non-vanishing order is
\be\n{rrr}
{\cal R}^2={48 M^2\over r^6}\, ,
\ee
as it must be for the Schwarzschild metric. Let us emphasize that
the same result can be obtained directly from the perturbation of the metric in the 
form (\ref{dhr}).

\subsection{Ghost-free case}

One can easily repeat similar calculations for the an arbitrary function
$F(z)$, where where $z=r^2-t^2$. We used the GRTensor program for this purpose. The 
calculations are
straightforward. However, the intermediate formulas are quite long. That is why we 
do not reproduce
them here. Let us only present the expression for the Kretschmann invariant in its 
lowest order for
such the metric in the ghost-free gravity
\be
{\cal R}^2={48 M^2 z^2\over r^6}Q(z)\hhh
Q(z)= 2{{\cal F}'}^2 z^2+2 {\cal F} {{\cal F}'} z+{\cal F}^2\hhh
{\cal F}=F' \, .
\ee
For the ghost-free theory with $a(\Box)=\exp(-\Box/\mu^2)$ one has
\ba
F(z)&=&Ei(1,z/(4s))+\gamma+\ln(z/(4s))\, ,\\
{\cal F}(z)&=&{1\over z}\left( 1-e^{-z/4s}\right)\, ,\\
Q(z)&=& {1\over 8 s^2 z^2}\left( (8s^2+4sz+z^2)e^{-z/2s}
-4s(4s+z) e^{-z/4s}+8s^2\right)\, .
\ea
Using the expansion (\ref{Fexp}) of $F$ for small $z$
\be
F(z)={z\over 4s}-{z^2\over 64s^2}+\ldots\, .
\ee
one finds
\be
{\cal R}^2={3M^2 z^2\over s^2 r^6}+\ldots= {3M^2 \beta^2\over s^2 r^4}+\ldots\, .
\ee
Here $\beta^2=1-t^2/r^2$ is a dimensionless parameter which outside the shell, where 
$|t|<r$, is
less or equal to 1. This means that the curvature for the linearized ghost-free 
theory in a case of
the collapse of the spherical null shell is weaker than the singularity for the 
linearized Einstein
equations. However, the ghost-free gravity solution still  remains singular at least 
for the chosen
scheme of the regularization.

\section{Spherical "thick" null shell collapse: Results}

\subsection{"Thick" null shell model}

In the previous section we discussed the gravitational field of a spherical null 
thin shell. This
shell represents a spherical $\delta$-type distribution of the energy. The mass of 
the shell is $M$.
It collapses with the speed of light  and  shrinks to zero radius at the moment of 
time $t=0$. In
the linearized theory the corresponding perturbation of the background flat metric 
is $\langle
dh^2\rangle(t,r)$. Certainly, such a model is an idealization. To study more 
realistic model of the
collapse we assume that the spherical collapsing null fluid is represented by a 
pulse, which is not
infinitely sharp in time, but has final time duration. We characterize its profile 
by a function
$q(t)$. The meaning of this function is the mass density per a unit time, $dM/dt$, 
arriving to the
center $r=0$ at time $t$. The total mass of such "thick" shell is \be
M=\int dt\  q(t)\, .
\ee
Using the linearity of the equations one can write the corresponding solution $\llan 
dh^2\rran (t,r)$  as the superposition of $\langle dh^2\rangle(t,r)$  perturbations 
as follows
\be\n{llrr}
\llan dh^2\rran (t,r)=\int dt' q(t') \langle dh^2\rangle(t-t',r)\, .
\ee

\begin{figure}[tbp]
\centering
\includegraphics[width=7cm]{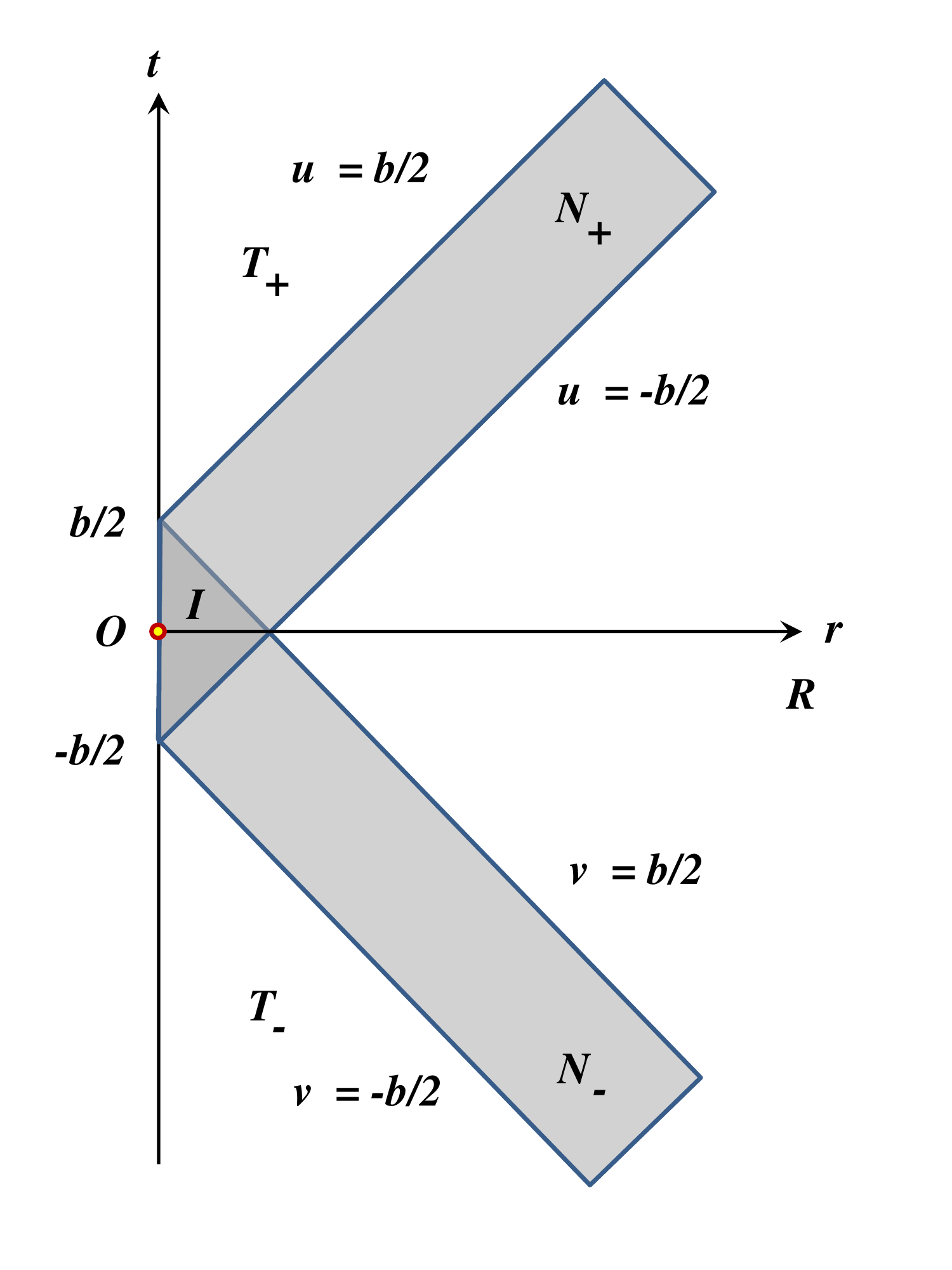}\hfill
  \caption{Thick null shell collapse\label{Fig_2}}
\end{figure}

To simplify the calculations we further specify the model. Namely, we choose $q(t)$ 
to be a step
function, which has a constant value $M/b$ during the interval $t\in(-b/2,b/2)$, and 
which vanishes
outside this interval. For such a step function a solution has different form in the 
different
spacetime domains (see Figure~\ref{Fig_2}). Let us describe first these domains. It 
is convenient to
use the advanced time, $v=t+r$, and the retarded time, $u=t-r$, coordinates. In the 
domains $T_-$,
where $v<-b/2$, and $T_+$, where $u>b/2$, the spacetime is empty, and the  
metric is flat there. In
the domain $N_-$, where $v\in(-b/2,b/2)$ and $u<-b/2$, one has only in-falling null 
fluid flux,
while in $N_+$, where $u\in (-b/2,b/2)$ and $v>b/2$ one has only out-going null 
fluid flux. In the
domain, $I$, where $v\in(-b/2,b/2)$ and $u\in (-b/2,b/2)$ one has a superposition of 
the in-coming
and out-going null fluid fluxes. And finally, in the domain $R$, where $v>b/2$ and 
$u<-b/2$, the
spacetime is empty.

\subsection{Gravitational field in the $I$-domain}

Let us consider first $I$-domain. Denote by $(t,r)$ the coordinates of the point of 
"observation" in
this domain. A thin null shell, crossing the center $r=0$ at time $t'$, contributes 
to the integral
(\ref{llrr}) only if $t'\in(t-r,t+r)$. Let us denote $x=t'-t$, then the formula 
(\ref{llrr}) takes
the form \ba
&&\llan dh^2\rran (t,r)=-{2M\over b r}\int_{-r}^{r} dx  H(r,x)\, ,\n{H1}\\
&&H(r,x)= F(r^2-x^2)\left( dt^2 -2{x\over r}dr dt +{x^2\over r^2} dr^2 +{1\over 
2}(r^2-x^2)
d\omega^2\right)\, .\n{H22}
\ea
Thus one has
\be
\llan dh^2\rran (t,r)=-{2M\over b r}\left[ J_0 (dt^2 +{1\over 2}r^2 d\omega^2)+J_2 
({dr^2\over r^2}-{1\over 2} d\omega^2)\right]\, .
\ee
Here
\be
J_n=\int_{-r}^{r} dx \ x^n\  F(r^2-x^2)\, .
\ee
The linear in $x$ term in (\ref{H2}) disappears as a result of the integration.

\subsubsection{Linearized Einstein theory}

For this case
\ba
&&F(z)=\ln z +C\hh C=-\ln\eta^2\, ,\\
&&J_0=4r(\ln 2+\ln r-1)+2C r\, ,\\
&&J_2={4\over 9}r^3 (3\ln 2+3 \ln r-4)+{2\over 3}Cr^3\, .
\ea
Calculations give
\be
{\cal R}^2={252 \dot{M}^2\over r^4}\hh \dot{M}=M/b\, .
\ee

\subsubsection{Ghost-free gravity case}

One has
\be
F(z)=\mbox{Ei}(1,z/4s)+\gamma +\ln(z/4s)\, .
\ee
The Tailor expansion of this function for small $z$ is
\be
F(z)={z\over 4s}-{z^2\over 64 s^2}+{z^2\over 1152 s^3}+\ldots\, .
\ee
Using this expansion one gets for small $r$ in the $I$-domain
\ba
J_0&=&{r^3(420 s^2-21s r^2+r^4)\over 1260 s^3}+\ldots\, ,\\
J_2&=&{r^5(756 s^2-27s r^2+r^4)\over 11340 s^3}+\ldots\,
\ea
Calculations of the Kretschmann invariant ${\cal R}^2$ for small $r$ in the 
$I$-domain give
\be
{\cal R}^2={32 \dot{M}^2\over 3 s^2}-{32 \dot{M}^2 r^2\over 9 s^3}+\ldots \, .
\ee
Here, as earlier, $\dot{M}=M/b$. This relation shows that for the thick shell model 
the curvature
remains finite at $r=0$.
\subsubsection{$(\nabla r)^2$ invariant}

There exists another useful invariant for the spherically symmetric geometry. This 
invariant is
\be
(\nabla r)^2={1\over 4 f} g^{\mu\nu} f_{,\mu} f_{,\nu}\hh
f=g_{\theta\theta}\, .
\ee
A line in the $(t,r)$ plane, where this invariant vanishes, is an apparent horizon. 
Using GRTensor
one find that in the leading order in $M$ this invariant in the $I$-domain is \be
(\nabla r)^2=1-{2M r^2\over s b}+\ldots\, .
\ee
Since in this domain $r<b$, one has
\be
(\nabla r)^2> 1-{2Mb\over s}\, .
\ee
This relation means that for given $s$, which is the square of the UV cut-off 
parameter $\mu$ of the
ghost-free theory, $s=\mu^{-2}$, and fixed duration $b$ of the pulse the mini-black 
hole is not
formed
if the mass $M$ is small enough.

\subsection{Gravitational field in the $R$ domain}

The perturbation of the gravitational field in the $R$ domain is described by the 
following expression
\be\n{J123a}
\llan dh^2\rran (t,r)=-{2M\over b r}\left[ J_0 (dt^2 +{1\over 2}r^2 d\omega^2)-J_1
{2 dt\ dr\over r}+J_2 ({dr^2\over r^2}-{1\over 2} d\omega^2)\right]\, , \ee
where now
\be\n{JJJ}
J_n=\int_{t-b/2}^{t+b/2} dx \ x^n\  F(r^2-x^2)\, .
\ee
Here
\ba
&&F(z)=F_0(z)+\Delta F(z)\hhh
F_0(z)=\gamma +\ln(z/4s)\, ,\\
&&\Delta F(z)=\mbox{Ei}(1,z/4s)\, .
\ea
We denote by $J^0_n$ and $\Delta J_n$ the contribution to $J_n$ of the terms $F_0$ 
and $\Delta F$,
respectively, so that one has \be
J_n=J^0_n+\Delta J_n\, .
\ee

We consider the perturbed metric at large $r$ in the $R$ domain. We also assume that 
$r\gg |t\pm
b/2|$ and use the following asymptotic form of the function $\Delta F(z)$ for large 
$z$ \be
\Delta F(z)={4s\over z}e^{-z/4s} (1+O(1/z))\, .
\ee
Thus one can use the following approximation
\be
\Delta F(r^2-x^2)\approx f(r)k(x)\hhh
f(r)={1\over r^2}e^{-r^2/(4s)}\hhh
k(x)=4s e^{x^2/(4s)}\, .
\ee
Using this approximation and (\ref{JJJ}) one obtains
\be
\Delta J_n=f(r) A_n(t)\hhh
A_n(t)=(4s)^{ {n+2\over 2}}\int_{C_-}^{C_+} dy \, y^n\, e^{y^2}\hhh
C_{\pm}={t\pm b/2\over 2\sqrt{s}}\, .
\ee
Calculation of $J_n^0$ is straightforward and $J^0_n$ can be easily obtained by 
using the Maple
program. One also obtains the following expressions for $A_n(t)$ \ba
&&A_0(t)=-4is^{3/2} \sqrt{\pi} \left ( 
\mbox{erf}(i\tau_+)+\mbox{erf}(i\tau_-)\right)\, ,\\
&&A_1(t)=8s^2\left( e^{\tau_+^2}-e^{\tau_-^2}\right)\, ,\\
&&A_3(t)=4s^2\left[ 2i\sqrt{\pi s}\left( 
\mbox{erf}(i\tau_+)+\mbox{erf}(i\tau_-)\right)+
(2t+b) e^{\tau_+^2}-(2t-b)e^{\tau_-^2}\right]\, .
\ea
Here
\be
\tau_{\pm}={\pm 2 t+b\over 4\sqrt{s}}\, ,
\ee
and $\mbox{erf}$ is the Error function, which is defined as
\be
\mbox{erf}(x) = {2\over \sqrt{\pi}}\int_0^x dt \exp(-t^2)\, .
\ee
Let us emphasize that in spite of the presence of the imaginary unit $i$ in the 
above formulas,
expressions for $A_n(t)$ are real, as they should be. After straightforward 
calculations by using
the GRTensor program, we obtain the following expression for the Kretschmann tensor 
in the leading
$M^2$ order
\ba\n{asRR}
&& {\cal R}^2={48 M^2\over r^6}+\Delta {\cal R}^2\hh
\Delta {\cal R}^2={2 M^2\over s^2 b r^{10}}\ e^{-r^2/( 4s)}\,  W
\, ,\\
&& W=-3 A_0 r^6+ (4\ddot{A}_0 s^2 +8 \dot{A}_1 s -34 A_0 s+A_2)r^4\nonumber\\
&&+(80 \dot{A}_1 s^2 -12 \ddot{A}_2 s^2 -160 A_0 s^2+14 A_2 s)r^2
+56 A_2 s^2\, .
\ea
Keeping the first term in $W$, which contains the highest power of $r$, we obtain
\be\n{DRRR}
\Delta {\cal R}^2\sim {24 i\sqrt{\pi} M^2\over \sqrt{s} b r^4}\ e^{-r^2/( 4s)}\,
\left( \mbox{erf}(i\tau_+)+\mbox{erf}(i\tau_-)\right)\, .
\ee
Once again, in spite of the presence of $i$ the answer for $\Delta {\cal R}^2$ is 
real.

Expressions (\ref{asRR}) and (\ref{DRRR}) imply that in the $R$ domain for a fixed 
value of time $t$ and $r\to \infty$ the curvature exponentially fast reaches its 
asymptotic value, which is equal to the Schwarzschild curvature of a spherical 
object of mass $M$.

\subsection{Gravitational field in the $N_{\pm}$-domains}

Let us briefly discuss now the gravitational field in the $N_{\pm}$ domains. We 
focus on the
incoming
$N_-$ domains. The case of the $N_+$ is similar. Let $(t,r)$ be coordinates of the 
"observation"
point. We denote $v=t+r$ and $u=t-r$. The condition that the "observation" point is 
in the $N_-$
domain reads \be
v\in (-b/2,b/2)\hh u\in (-\infty,-b/2)\, .
\ee

The position $t'$ of the apex of the null-shells, that give non-zero contribution at 
the point of
the "observation", obeys the conditions $t'\in (-b/2,t+r)$. As earlier, we denote 
$x=t-t'$. Then \ba
&&\llan dh^2\rran (t,r)=-{2M\over b r}\int_{-r}^{t+b/2} dx  H(r,x)\, , \\
&&H(r,x)= F(r^2-x^2)\left( (dt-{x\over r}dr)^2 +{1\over 2}(r^2-x^2) 
d\omega^2\right)\, .\n{H2}
\ea
Thus one has
\be\n{J123}
\llan dh^2\rran (t,r)=-{2M\over b r}\left[ J_0 (dt^2 +{1\over 2}r^2 d\omega^2)-J_1
{2 dt\ dr\over r}+J_2 ({dr^2\over r^2}-{1\over 2} d\omega^2)\right]\, . \ee
Here
\be
J_n=\int_{-r}^{t+b/2} dx \ x^n\  F(r^2-x^2)\, .
\ee

In the linearized Einstein gravity one uses the following expression for $F$
\be
F(r^2-t^2)=\ln\left({r^2-t^2\over \eta^2}\right)\, .
\ee
The required integrals (\ref{JJJ}) can be easily calculated. We do not reproduce 
here these results
since the obtained expressions are quite long. We present here only final expression 
for the
Kretschmann tensor, which we obtained by using th GRTensor program \be
{\cal R}^2={48 M^2(v)\over r^6}\hh M(v)=\dot{M} (v+b/2)
\, .
\ee
This is an expected result. Indeed, before the collapsing null fluid meets the 
outgoing flux, that
is in the $N_-$-domain, one has simply the Vaidya solution with mass $M(v)={M\over 
b} (v+b/2)$,
where $v\in (-b/2,b/2)$. For a chosen model $M(v)$ linearly grows from zero (at 
$v=-b/2$) till its
maximal value $M$.
Calculations for the metric perturbation in the $N_+$ are similar and give the 
following result
\be
{\cal R}^2={48 M^2(u)\over r^6}\hh M(u)=\dot{M} (u+b/2)\, .
\ee

In the case of the ghost-free gravity the curvature is modified in the narrow "skin" 
domains close
to $v=\pm b/2$ (for $N_-$ domain) and close to $u=\pm b/2$ (for $N_+$ domain), while 
outside of them
and at large $r$ the ghost-free theory corrections only slightly modify the above 
described Vaidya
metric.

\section{Summary and discussion}

Let us summarize and discuss the obtained results. We remind that we study the 
modification of the
solutions of the Einstein theory in the framework of the ghost-free theory, proposed 
in
\cite{Biswas:2011ar,Modesto:2011kw,Biswas:2013cha,Biswas:2013kla}. We focus mainly 
on a special 
type of such a
theory in which a free field operator $\Box$ is  replaced by the non-local 
operator of the form
$\exp(-\Box/\mu^2)\,\Box$. We study solutions of such a theory in the linearized 
approximation. We
first study a gravitational field of a point mass in this theory. Such a problem was 
solved in the
four dimensional space time in \cite{Biswas:2011ar,Modesto:2010uh} by using 
the Fourier methods. We demonstrate
that the propagator of the linearized ghost-free theory in any number of spacetime 
dimensions is
directly related to the heat kernel in this space. Using the heat kernel approach we 
obtain
solutions for the gravitational field of a point mass in $D$-dimensional spacetime. 
We demonstrate
that these solutions are always regular at the position of the source. In $4D$  
spacetime the
obtained solution coincides with the result presented in 
\cite{Biswas:2011ar,Modesto:2010uh}.

In the second part of the paper we study solutions of
the ghost-free gravity for ultra-relativistic sources. We first obtain ghost-free 
analogues of
gyraton solutions \cite{Frolov:2005in,Frolov:2005zq,FrolovZelnikov:2011} in the 
ghost-free
gravity. For this purpose we boost the obtained ghost-free solution for a point 
mass, and by taking
the Penrose limit, we found a  solution for a $D$-dimensional spinless ``photon". We
demonstrate that the corresponding metric is similar to the metric of a spinless 
gyraton. The main
difference is that a solution of $(D-2)$-dimensional flat Laplace equation for a 
point charge, that
enters the gyraton metric, is modified. Namely, the  function of the transverse 
variables becomes a
solution of $(D-2)$ ghost-free modification of the corresponding Laplace operator. 
Using this result
we obtain  spinless gyraton solutions of the ghost-free gravity in $D$-dimensional 
spacetime and
demonstrate that their transverse singularities are regularized.

Finally, we study a spherical gravitational collapse of null fluid in the framework 
of the
ghost-free
gravity. As earlier, we restrict ourselves by working in the linearized 
approximation. Since a
solution of the ghost-free equations for the dynamical problem seems to be 
complicated, we
used the following trick based on the results, obtained earlier in this paper. We 
used a fact that
in the linearized theory a superposition of any solutions is again a solution. To 
obtain the
gravitational field for a collapsing spherical thin null, we "construct" it as a 
superposition of
the gyraton metrics for a spherically symmetric distribution of the gyraton sources 
moving along a
null cone, representing the shell. Such gyratons intersect at a single point of the 
spacetime, which
is an apex of the null cone. In the adopted linearized approximation the gyratons, 
forming the thin
shell, cross this point simultaneously without interaction, so that after 
passing the apex point they form an
expanding spherical null shell. The gravitational field for such null shells is 
obtained by
averaging of the single gyraton metric over homogeneous spherical distribution of 
the gyratons. We
demonstrated that by using this approach one correctly reproduces the stress-energy 
tensor for
both,
collapsing and expanding thin null shells. We checked that the obtained solution in 
the
linearized Einstein gravity correctly reproduces the expected result. Namely, the 
gravitational
field
inside both, collapsing and expanding shells vanishes, while outside of them it is 
time independent. The
latter field is nothing but a linearized version of the Schwarzschild metric with 
mass $M$. After
this, we obtained a similar solution for the ghost-free gravity. To study its 
properties we
calculated the Kretschmann scalar for it and demonstrated that its singularity at 
$r=0$  is
smoothened,
but still is present.

The model of an infinitely thin null shell is certainly an idealization. One 
can expect that in a consistent theory the non-locality cannot be a property of the
gravitational field only, but similar non-locality must be present in the description of 
matter and other fields. For this reason a ``physical" collapsing shell must have 
natural finite thickness, which may regularize the curvature singularity of the thin 
shell. In order to check this assumption, we study a case of the thick null shell 
collapse.
In the linearized theory a corresponding
solution is obtained by a superposition of the thin-shell solutions, that is by 
averaging these
solutions with different positions of their apexes at $r=0$. As a result, we 
construct a solution
for
a thick null shell with an arbitrary distribution of the collapsing mass $M=M(v)$ at 
the spatial
infinity. We considered in detail a special model, where the mass $M(v)$ is a linear 
function of
the
advanced time $v=t+r$, such that $\dot{M}$ vanishes for the moments $v<-b/2$ and 
$v>b/2$, and remains 
constant inside
this interval. We demonstrated, that in the linearized Einstein gravity the solution 
is a
superposition of two linearized Vaidya metrics, for the incoming and outgoing null 
fluid fluxes. The
corresponding Kretschmann scalar is ${\cal R}^2={252 \dot{M}^2\over r^4}$, where 
$\dot{M}=M/b$. It
is singular at $r=0$.

In the linearized ghost-free gravity the corresponding solution is modified. The 
main new feature
is, that the metric is regular at $r=0$. As a result, its Kretschmann scalar ${\cal 
R}^2={32
\dot{M}^2\over 3 s^2}$ is finite at $r=0$. We also calculated the invariant $(\nabla 
r)^2$ and
demonstrated for the collapse of a thick null shell of the small mass $M$ the 
apparent horizon is
not formed. This result can be interpreted as follows: in the ghost-free gravity 
there exists a mass
gap for the black hole formation in the gravitational collapse. This result is 
similar to the result
obtained in \cite{Frolov:1981mz} for the null shell collapse in the theory gravity 
with quadratic in
the curvature corrections.
Recently  it was demonstrated that the mass gap for mini black hole formation is a 
common property not only of the ghost-free gravity, but also of a wide class of higher 
derivative theories of gravity \cite{Frolov:2015bta}. These theories contain a mass 
scale parameter $\mu$, which plays the role of the ultra-violet cut-off. As a result, 
if the mass $M$ of a collapsing object obeys the relation $M\mu\lesssim 1$, an apparent 
horizon is not formed. The presence of such a scale parameter, differs these 
theories from the classical Einstein gravity, where
the mass gap is absent and the black-holes of arbitrary small mass can be formed. A 
well known
consequence of this is Choptuik \cite{Choptuik:1992jv} type universal scaling 
properties of
near-critical solutions.

Let us make a few general remarks concerning solutions of the ghost-free gravity 
equations. The heat kernels, which we used to construct solutions for a static 
source, are well defined in the space
with the Euclidean metric. The reason is that the Laplace operator is non-positive 
definite. However this property is not valid for the box operator. In this paper we 
used the method based on the gyraton solutions to overcome
this problem for a special case of the null shell collapse. It would be interesting 
to investigate solutions of
the linearized ghost-free gravity with the form-factor $a=\exp(-\Box/\mu^2)$ for 
arbitrary moving sources, say, for the emission of the
gravitational waves in such a theory, and the back-reaction of this radiation on the 
accelerated
objects. Another proposed option is to modify this form-factor, and instead of 
$\Box$ to consider higher powers of this operator in the exponent. A simplest 
example is $a=\exp(\alpha\Box^2)$. It will be interesting to study such 
modifications.

In the present paper we study solutions of the linearized equations of the ghost 
free theory. Let us emphasize, that this  is sufficient for demonstration of the 
existence of the mass gap for mini black hole formation in the adopted model. The 
reason is that if such a solution for small mass $M$ is regular and the 
corresponding perturbation of the flat metric is uniformly small, then one can 
expect that the higher order corrections to this solution can be neglected. A 
natural question is what happens in the collapse of large mass $M$, that is when 
$M\mu\gg 1$. It would be highly interesting to analyze solutions of the ghost-free 
gravity in this regime \cite{Modesto:2010uh,Bambi:2013gva}. It is natural to assume  
that
curvature remains finite inside the black holes and the limiting curvature 
conjecture is satisfied \cite{Markov:1982,Markov:1984ii,Polchinski:1989ae}. These was 
a lot of discussions of non-singular models of black holes. One of the option is that 
the apparent horizon is closed \cite{Frolov:1981mz} (see also 
\cite{Hayward:2005gi,Frolov:2014jva,Bardeen:2014uaa} and references therein). Another 
option is a new universe formation inside a black hole 
\cite{Frolov:1989pf,Frolov:1988vj}. This 
option was also widely discussed in the literature. One may hope that proposed ghost 
free modifications of the Einstein gravity, which are ultraviolet complete and 
asymptotically free, would allow one to answer  intriguing questions concerting the 
structure of the black hole interior.

\vspace{1.5cm}

\section*{Acknowledgments}

The authors thank the Natural Sciences and Engineering Research Council of Canada 
for the financial
support. Two of the authors (V.F. and A.Z.) are also grateful to the Killam Trust 
for its financial
support. T.P.N. is grateful to CAPES and Natural Sciences and Engineering Research 
Council of Canada
for supporting his visit to the University of
Alberta.

\appendix

\section{Useful geometric relations}

\begin{figure}[tbp]
\centering
\includegraphics[width=13cm]{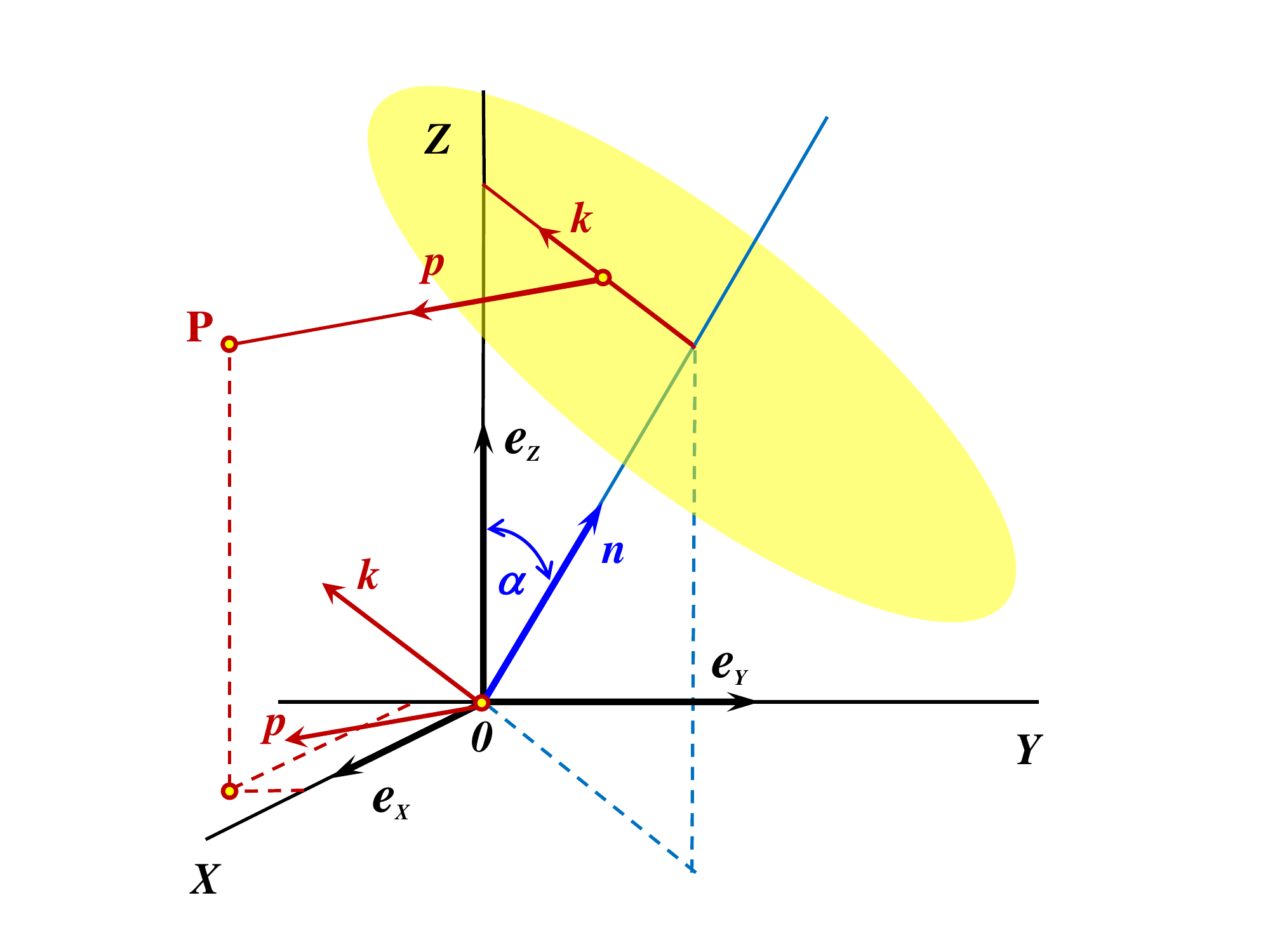}\hfill
  \caption{Relation between the Cartesian and the gyraton based 
coordinates\label{Fig_1}}
\end{figure}

In order to find this coordinate transformation let us introduce a couple of new 
useful unit
vectors. The first one is a vector $\BM{k}$ which is orthogonal to $\BM{n}$ and is 
directed from the
gyraton trajectory to the $Z$-axis (see Figure~\ref{Fig_1}). It can be written as a 
linear
combination \be
\BM{k}=k_{\inds{Z}} \BM{e}_{\inds{Z}}+k_{\inds{n}} \BM{n}\, .
\ee
Using relation $(\BM{n},\BM{e}_{\inds{Z}})=\cos\alpha$ one finds
\be\n{kk}
\BM{k}={1\over \sin\alpha}( \BM{e}_{\inds{Z}}-\cos\alpha \BM{n})\, .
\ee
The second unit vector $\BM{p}$ is orthogonal to both $\BM{n}$ and $\BM{k}$. It can 
be written as
\be
\BM{p}=[\BM{n}\times \BM{k}]={1\over \sin\alpha}[\BM{n}\times \BM{e}_{\inds{Z}}]\, .
\ee
Hence this vector is orthogonal to $\BM{e}_{\inds{Z}}$ and lies in the $X-Y$-plane. 
It is easy to
check that \be\n{pp}
\BM{p}=\sin\beta\, \BM{e}_{\inds{X}} -\cos\beta \, \BM{e}_{\inds{Y}}\, .
\ee

Consider a point $P$, which has the Cartesian coordinates $(X,Y,Z)$, so that the 
vector $\BM{P}$
connecting the origin point $O$ with $P$ is \be\n{ccar}
\BM{P}=X\BM{e}_{\inds{X}}+Y\BM{e}_{\inds{Y}}+Z\BM{e}_{\inds{Z}}\, .
\ee
The same vector can be written in the gyraton's frame as follows 
($\zeta_{\inds{\perp}}=(\zeta_k,\zeta_p)$)
\be\n{cgyr}
\BM{P}=\xi \BM{n}+\zeta_k \BM{k}+\zeta_p \BM{p}\, .
\ee
By comparing (\ref{ccar}) and (\ref{cgyr}) and using relations (\ref{nn}), 
(\ref{kk}) and (\ref{pp}) one finds
\ba
X&=&\cos\beta \, (\xi \sin\alpha -\zeta_k \cos\alpha)+\zeta_p\, \sin\beta\, 
,\nonumber\\
Y&=&\sin\beta\, (\xi \sin\alpha -\zeta_k \cos\alpha)-\zeta_p \, \cos\beta\, ,\\
Z&=&\xi\, \cos\alpha +\zeta_k\, \sin\alpha\, .\nonumber
\ea
We shall also use the following inverse relations
\ba
\xi&=&\sin\alpha\, (\cos\beta X+\sin\beta Y)+\cos\alpha\, Z\, ,\n{xixi}\\
\zeta_k&=&-\cos\alpha \, (\cos\beta X+\sin\beta Y)+\sin\alpha \, Z\, ,\n{zek}\\
\zeta_p&=&\sin\beta\, X-\cos\beta\, Y\, .\n{zep}
\ea
Using these results it is easy to show that
\be
d\xi^2+d\zeta_{\inds{\perp}}^2=
dX^2+dY^2+dZ^2\, .
\ee

\section{Evaluating integrals}

We assume that $\zeta_p$ takes the value $y_p$ and use relation (\ref{zep}) to find 
a corresponding
value of the angle $\beta$. By solving this equation one gets \ba\n{scb}
&&\sin(\beta_{\pm})={y_p X +Y Q_{\pm}\over X^2+Y^2}\, ,\\
&&\cos(\beta_{\pm})={-y_p Y +X Q_{\pm}\over X^2+Y^2}\, ,\\ 
&&Q_{\pm}=\pm\sqrt{X^2+Y^2-y_p^2}\, .
\ea
These relations show that for each value of $y_p$ there exist two different values 
of the angle
$\beta$, which we denoted by $\beta_{\pm}$. Using (\ref{scb}) it is easy to check 
that \be
\cos\beta_{\pm} X +\sin\beta_{\pm} Y= Q_{\pm}\, ,
\ee
and the relations (\ref{xixi}) and (\ref{zek}) take the form
\ba
\xi&=&\sin\alpha Q_{\pm}+\cos\alpha Z\, ,\n{xx}\\
\zeta_{k}&=&-\cos\alpha Q_{\pm}+\sin\alpha Z\,.\n{zz}
\ea

Let us calculate the following integral
\be\n{bb}
[\ldots]_{y_p}={1\over 2\pi}\int_0^{2\pi} d\beta \, \delta(y_p-\zeta_p)\, (\ldots)\, 
.
\ee
Using the relation (\ref{zep}) one gets
\be\n{zb}
d\zeta_p=|\partial_{\beta}\zeta_p|\, d\beta =Q \, d\beta\hh Q=Q_{+}\, .
\ee
Using this relation one rewrites (\ref{bb}) as an integral over $\zeta_p$ which can 
be easily taken
with the result \be
[\ldots]_{y_p}={1\over 2\pi Q}\left[ (\ldots)_{y_p}^+ + (\ldots)_{y_p}^-\right]\, .
\ee
The expression in the right-hand side is a sum of the contributions for two values 
of $\beta_{\pm}$
corresponding to the same $y_p$. The quantity $\zeta_p$ in each of these terms 
should be substituted
by $y_p$.
We use a similar trick as earlier to calculate an integral over the angle variable 
$\alpha$. First,
we use the relation (\ref{zek}) with $\zeta_k=y_k$ to find the angle $\alpha$. One 
gets \ba
\sin\alpha_{\pm}&=&{y_k Z+Q_+\sqrt{r^2-y_{\inds{\perp}}^2}\over r^2-y_p^2}\, 
,\n{sa}\\
\cos\alpha_{\pm}&=&{-y_k Q_{\pm}\pm Z\sqrt{r^2-y_{\inds{\perp}}^2}\over r^2-y_p^2}\, 
,\n{ca}\\
r^2&=&X^2+Y^2+Z^2\hh y_{\inds{\perp}}^2=y_p^2+y_k^2\, .
\ea
One also has
\be\n{xxx}
\xi_{\pm}=\pm\sqrt{r^2-y_{\inds{\perp}}^2}\, .
\ee

In order to obtain this results one needs to solve  quadratic equations, which have 
two solutions.
We  single out a solution as follows. A line $y_k=y_p=0$ coincides with the 
gyraton's trajectory and
$\alpha$ and $\beta$ are spherical angles of its direction. For $y_k=y_p=0$ 
relations (\ref{sa}),
(\ref{ca}) and (\ref{xxx}) give \be\n{scay}
\sin\alpha_{\pm}={\sqrt{X^2+Y^2}\over r}\hhh
\cos\alpha_{\pm}=\pm{Z\over r}\hhh \xi_{\pm}=\pm r
\, .
\ee
Namely these conditions (\ref{scay}) fix an ambiguity in the sign choice in the 
relations (\ref{sa}) and (\ref{ca}).
Let us notice also that (\ref{scb}) in the limit $y_k=y_p=0$  takes the form
\be\n{scby}
\sin(\beta_{\pm})=\pm{Y \over \sqrt{X^2+Y^2}}\hh
\cos(\beta_{\pm})=\pm{X \over \sqrt{X^2+Y^2}}\, .
\ee
By using (\ref{scay}) and (\ref{scby}) it is easy to see that two solutions 
$(\alpha_+,\beta_+)$ and
$(\alpha_-,\beta_-)$ describe two opposite points of a unit sphere. As we explained 
earlier these
two solutions describe two gyraton trajectories passing through the same space point 
$(X,Y,Z)$.

Using relations (\ref{xx}) and (\ref{zz}) one gets
\be
\partial_{\alpha}\zeta_{k}=\xi\hh  d\zeta_k=|\xi| \, 
d\alpha=\sqrt{r^2-y_{\inds{\perp}}^2}\, d\alpha\, .
\ee
Let us denote
\be
\langle \ldots\rangle_{y_{\inds{\perp}}}={1\over 2}\int_0^{\pi} d\alpha\, \sin\alpha\
\delta(y_k-\zeta_k) \  [\ldots]_{y_p}\, , \ee
then one has
\be\n{aver}
\langle \ldots\rangle_{y_{\inds{\perp}}}={1\over 4\pi Q 
\sqrt{r^2-y_{\inds{\perp}}^2}}\left[
\sin\alpha_+(\ldots)^+ + 
\sin\alpha_-(\ldots)^-\right]_{\zeta_{\inds{\perp}}=y_\inds{\perp}}\, . 
\ee

\providecommand{\href}[2]{#2}\begingroup\raggedright\endgroup


\begin{thebibliography}{10}

\bibitem{Myrzakulov:2013hca}
R.~Myrzakulov, L.~Sebastiani, and S.~Zerbini, {\it {Some aspects of generalized
  modified gravity models}},  {\em Int.J.Mod.Phys.} {\bf D22} (2013) 1330017,
  [\href{http://xxx.lanl.gov/abs/1302.4646}{{\tt arXiv:1302.4646}}].

\bibitem{Stelle:1977ry}
K.~S. Stelle, {\it {Classical Gravity with Higher Derivatives}},  {\em
  Gen.Rel.Grav.} {\bf 9} (1978) 353--371.

\bibitem{Frolov:1981mz}
V.~P. Frolov and G.~Vilkovisky, {\it {Spherically Symmetric Collapse in Quantum
  Gravity}},  {\em Phys.Lett.} {\bf B106} (1981) 307--313.

\bibitem{Modesto:2014eta}
L.~Modesto, T.~d.~P. Netto, and I.~L. Shapiro, {\it {On Newtonian singularities
  in higher derivative gravity models}},
  \href{http://xxx.lanl.gov/abs/1412.0740}{{\tt arXiv:1412.0740}}.

\bibitem{Stelle:1976gc}
K.~Stelle, {\it {Renormalization of Higher Derivative Quantum Gravity}},  {\em
  Phys.Rev.} {\bf D16} (1977) 953--969.

\bibitem{Asorey:1996hz}
M.~Asorey, J.~Lopez, and I.~Shapiro, {\it {Some remarks on high derivative
  quantum gravity}},  {\em Int.J.Mod.Phys.} {\bf A12} (1997) 5711--5734,
  [\href{http://xxx.lanl.gov/abs/hep-th/9610006}{{\tt hep-th/9610006}}].

\bibitem{Biswas:2011ar}
T.~Biswas, E.~Gerwick, T.~Koivisto, and A.~Mazumdar, {\it {Towards singularity
  and ghost free theories of gravity}},  {\em Phys.Rev.Lett.} {\bf 108} (2012)
  031101, [\href{http://xxx.lanl.gov/abs/1110.5249}{{\tt arXiv:1110.5249}}].

\bibitem{Modesto:2011kw}
L.~Modesto, {\it {Super-renormalizable Quantum Gravity}},  {\em Phys.Rev.} {\bf
  D86} (2012) 044005, [\href{http://xxx.lanl.gov/abs/1107.2403}{{\tt
  arXiv:1107.2403}}].

\bibitem{Modesto:2012ys}
L.~Modesto, {\it {Super-Renormalizable Multidimensional Gravity: Theory and
  Applications}},  {\em Astron.Rev.} {\bf 8} (2012), no.~2 4--33,
  [\href{http://xxx.lanl.gov/abs/1202.3151}{{\tt arXiv:1202.3151}}].

\bibitem{Biswas:2013cha}
T.~Biswas, A.~Conroy, A.~S. Koshelev, and A.~Mazumdar, {\it {Generalized
  ghost-free quadratic curvature gravity}},  {\em Class.Quant.Grav.} {\bf 31}
  (2014) 015022, [\href{http://xxx.lanl.gov/abs/1308.2319}{{\tt
  arXiv:1308.2319}}].

\bibitem{Biswas:2013kla}
T.~Biswas, T.~Koivisto, and A.~Mazumdar, {\it {Nonlocal theories of gravity:
  the flat space propagator}},  \href{http://xxx.lanl.gov/abs/1302.0532}{{\tt
  arXiv:1302.0532}}.

\bibitem{Tomboulis:1997gg}
E.~Tomboulis, {\it {Superrenormalizable gauge and gravitational theories}},
  \href{http://xxx.lanl.gov/abs/hep-th/9702146}{{\tt hep-th/9702146}}.

\bibitem{Modesto:2014lga}
L.~Modesto and L.~Rachwal, {\it {Super-renormalizable and finite gravitational
  theories}},  {\em Nucl.Phys.} {\bf B889} (2014) 228--248,
  [\href{http://xxx.lanl.gov/abs/1407.8036}{{\tt arXiv:1407.8036}}].

\bibitem{Shapiro:2015uxa}
I.~L. Shapiro, {\it {Counting ghosts in the "ghost-free" non-local gravity}},
  {\em Phys.Lett.} {\bf B744} (2015) 67--73,
  [\href{http://xxx.lanl.gov/abs/1502.0010}{{\tt arXiv:1502.0010}}].

\bibitem{Modesto:2010uh}
L.~Modesto, J.~W. Moffat, and P.~Nicolini, {\it {Black holes in an ultraviolet
  complete quantum gravity}},  {\em Phys.Lett.} {\bf B695} (2011) 397--400,
  [\href{http://xxx.lanl.gov/abs/1010.0680}{{\tt arXiv:1010.0680}}].

\bibitem{Bambi:2013gva}
C.~Bambi, D.~Malafarina, and L.~Modesto, {\it {Terminating black holes in
  asymptotically free quantum gravity}},  {\em Eur.Phys.J.} {\bf C74} (2014)
  2767, [\href{http://xxx.lanl.gov/abs/1306.1668}{{\tt arXiv:1306.1668}}].

\bibitem{Biswas:2005qr}
T.~Biswas, A.~Mazumdar, and W.~Siegel, {\it {Bouncing universes in
  string-inspired gravity}},  {\em JCAP} {\bf 0603} (2006) 009,
  [\href{http://xxx.lanl.gov/abs/hep-th/0508194}{{\tt hep-th/0508194}}].

\bibitem{Khoury:2006fg}
J.~Khoury, {\it {Fading gravity and self-inflation}},  {\em Phys.Rev.} {\bf
  D76} (2007) 123513, [\href{http://xxx.lanl.gov/abs/hep-th/0612052}{{\tt
  hep-th/0612052}}].

\bibitem{Biswas:2010zk}
T.~Biswas, T.~Koivisto, and A.~Mazumdar, {\it {Towards a resolution of the
  cosmological singularity in non-local higher derivative theories of
  gravity}},  {\em JCAP} {\bf 1011} (2010) 008,
  [\href{http://xxx.lanl.gov/abs/1005.0590}{{\tt arXiv:1005.0590}}].

\bibitem{Barvinsky:2012ts}
A.~Barvinsky and Y.~Gusev, {\it {New representation of the nonlocal ghost-free
  gravity theory}},  {\em Phys.Part.Nucl.} {\bf 44} (2013) 213--219,
  [\href{http://xxx.lanl.gov/abs/1209.3062}{{\tt arXiv:1209.3062}}].

\bibitem{efimov1967}
G.~V. Efimov, {\it {Non-local quantum theory of the scalar field}},  {\em Comm.
  Math. Phys.} {\bf 5} (1967), no.~1 42--56.

\bibitem{Efimov:1972wj}
G.~Efimov, {\it {On the construction of nonlocal quantum electrodynamics}},
  {\em Annals Phys.} {\bf 71} (1972) 466--485.

\bibitem{Efimov:1976nu}
G.~Efimov, M.~A. Ivanov, and O.~Mogilevsky, {\it {Electron Selfenergy in
  Nonlocal Field Theory}},  {\em Annals Phys.} {\bf 103} (1977) 169--184.

\bibitem{Efimov:18}
G.~Efimov, {\it {Quantization of non-local field theory}},  {\em International
  Journal of Theoretical Physics} {\bf 10} (1974), no.~1 19--37.

\bibitem{Efimov:19}
G.~Efimov, {\it {On a class of relativistic invariant distributions}},  {\em
  Communications in Mathematical Physics} {\bf 7} (1968), no.~2 138--151.

\bibitem{Krasnikov:1987}
N.~Krasnikov, {\it {Nonlocal gauge theories}},  {\em Theoretical and
  Mathematical Physics} {\bf 73} (1987), no.~2 1184--1190.

\bibitem{Frolov:2005in}
V.~Frolov and D.~Fursaev, {\it {Gravitational field of a spinning radiation
  beam-pulse in higher dimensions}},  {\em Phys.Rev.} {\bf D71} (2005) 104034,
  [\href{http://xxx.lanl.gov/abs/hep-th/0504027}{{\tt hep-th/0504027}}].

\bibitem{Frolov:2005zq}
V.~P. Frolov, W.~Israel, and A.~Zelnikov, {\it {Gravitational field of
  relativistic gyratons}},  {\em Phys.Rev.} {\bf D72} (2005) 084031,
  [\href{http://xxx.lanl.gov/abs/hep-th/0506001}{{\tt hep-th/0506001}}].

\bibitem{FrolovZelnikov:2011}
V.~Frolov and A.~Zelnikov, {\em {Introduction to black hole physics}}.
\newblock Oxford University Press, 2011.

\bibitem{Misner:1974qy}
C.~W. Misner, K.~Thorne, and J.~Wheeler, {\em {Gravitation}}.
\newblock W.H. Freeman and Co., San Francisco, 1974.

\bibitem{Aichelburg:1970dh}
P.~Aichelburg and R.~Sexl, {\it {On the gravitational field of a massless
  particle}},  {\em Gen.Rel.Grav.} {\bf 2} (1971) 303--312.

\bibitem{Frolov:1998wf}
V.~Frolov and I.~Novikov, {\em {Black hole physics: Basic concepts and new
  developments}}.
\newblock Kluwer Acad. Publ., 1998.

\bibitem{Frolov:2015bta}
V.~P. Frolov, {\it {Mass-gap for black hole formation in higher derivative and
  ghost-free gravity}},  \href{http://xxx.lanl.gov/abs/1505.0049}{{\tt
  arXiv:1505.0049}}.

\bibitem{Choptuik:1992jv}
M.~W. Choptuik, {\it {Universality and scaling in gravitational collapse of a
  massless scalar field}},  {\em Phys.Rev.Lett.} {\bf 70} (1993) 9--12.

\bibitem{Markov:1982}
M.~Markov, {\it {Limiting density of matter as a universal law of nature}},
  {\em JETP Letters} {\bf 36} (1982) 266.

\bibitem{Markov:1984ii}
M.~Markov, {\it {Problems of a Perpetually Oscillating Universe}},  {\em Annals
  Phys.} {\bf 155} (1984) 333--357.

\bibitem{Polchinski:1989ae}
J.~Polchinski, {\it {Decoupling Versus Excluded Volume or Return of the Giant
  Wormholes}},  {\em Nucl.Phys.} {\bf B325} (1989) 619--630.

\bibitem{Hayward:2005gi}
S.~A. Hayward, {\it {Formation and evaporation of regular black holes}},  {\em
  Phys.Rev.Lett.} {\bf 96} (2006) 031103,
  [\href{http://xxx.lanl.gov/abs/gr-qc/0506126}{{\tt gr-qc/0506126}}].

\bibitem{Frolov:2014jva}
V.~P. Frolov, {\it {Information loss problem and a 'black hole` model with a
  closed apparent horizon}},  {\em JHEP} {\bf 1405} (2014) 049,
  [\href{http://xxx.lanl.gov/abs/1402.5446}{{\tt arXiv:1402.5446}}].

\bibitem{Bardeen:2014uaa}
J.~M. Bardeen, {\it {Black hole evaporation without an event horizon}},
  \href{http://xxx.lanl.gov/abs/1406.4098}{{\tt arXiv:1406.4098}}.

\bibitem{Frolov:1989pf}
V.~P. Frolov, M.~Markov, and V.~F. Mukhanov, {\it {Through a black hole into a
  new Universe?}},  {\em Phys.Lett.} {\bf B216} (1989) 272--276.

\bibitem{Frolov:1988vj}
V.~P. Frolov, M.~Markov, and V.~F. Mukhanov, {\it {Black holes as possible
  sources of closed and semiclosed worlds}},  {\em Phys.Rev.} {\bf D41} (1990)
  383--394.

\end{thebibliography}

\end{document}